\documentstyle[12pt,aaspp4]{article}
\def\icm{${\rm cm}^{-1}$}
\def\he3{$^3{\rm He}$}

\def\IRAS{{\sl IRAS}}
\def\uK{\hbox{$\mu$K}}

\begin{document}

% In version 4.0 aaspp style the \slugcomment must come before the
% \title or it won't appear.
\slugcomment{Submitted to {\em Ap. J. Letters} 18 August 1995}

\title{MSAM1-94: Repeated Measurement of Medium-Scale Anisotropy\\
in the Cosmic Microwave Background Radiation}

\author{
E.~S.~Cheng\altaffilmark{1},
D.~A.~Cottingham\altaffilmark{2},
D.~J.~Fixsen\altaffilmark{3},
C.~A.~Inman\altaffilmark{4},
M.~S.~Kowitt\altaffilmark{1},
S.~S.~Meyer\altaffilmark{4},
L.~A.~Page\altaffilmark{5},
J.~L.~Puchalla\altaffilmark{4},
J.~E.~Ruhl\altaffilmark{4},
and~R.~F.~Silverberg\altaffilmark{1}}

\altaffiltext{1}{NASA/Goddard Space Flight Center, Laboratory for Astronomy
and Solar Physics, Code 685.0, Greenbelt, MD 20771}
\altaffiltext{2}{Global Science and Technology, Inc., NASA/GSFC Laboratory
for Astronomy and Solar Physics, Code 685.0, Greenbelt, MD 20771}
\altaffiltext{3}{Applied Research Corporation, NASA/GSFC Laboratory for
Astronomy and Solar Physics, Code 685.3, Greenbelt, MD 20771}
\altaffiltext{4}{University of Chicago, 5640 S. Ellis St., Chicago, IL 
60637}
\altaffiltext{5}{Princeton University Physics Dept., Princeton, NJ 08544}

\begin{abstract}

The second flight of the Medium Scale Anisotropy Measurement (MSAM1-94)
observed the same field as
the first flight (MSAM1-92) to confirm our
earlier measurement of cosmic microwave background radiation (CMBR) anisotropy.
This instrument chops a 30\arcmin\ beam in a 3 position pattern with a
throw of $\pm40\arcmin$,
and simultaneously measures single and double differenced sky signals.
We observe in four spectral channels centered at 5.6,
9.0, 16.5, and 22.5~\icm, providing sensitivity to the peak of the
CMBR and to thermal emission from interstellar dust.
The dust component correlates well
with the \IRAS\ 100~\micron\ map.  The CMBR observations in our double
difference channel correlate well with the earlier observations, but
the single difference channel shows some discrepancies.
We obtain a detection of
fluctuations in the MSAM1-94 dataset that match CMBR
in our spectral bands of
$\Delta T/T = 1.9^{+1.3}_{-0.7}\times 10^{-5}$ (90\% confidence interval,
including calibration uncertainty)
for
total rms Gaussian fluctuations with correlation angle 0\fdg3, using
the double difference demodulation.

\end{abstract}

\section{Introduction}

Observations of anisotropy in the Cosmic Microwave Background
Radiation (CMBR) yield valuable clues about the formation of
large-scale structure in the early universe.
A particularly interesting angular scale for
observing CMBR anisotropy is near 0\fdg5, where the first ``Doppler peak''
(or adiabatic peak)
enhancement of the fluctuation power spectrum
is expected to be observable (\cite{white94}).
The Medium Scale Anisotropy
Measurement (MSAM) is an experiment designed to measure CMBR anisotropy at
this angular scale.  This paper reports the initial results from the
second flight of this experiment.

A number of detections of anisotropy at angular scales near 0\fdg5
have been reported recently.  Observations by ARGO
(\cite{debernardis94}), the Python experiment (\cite{dragovan94}), the
fourth flight of the MAX experiment (\cite{devlin94,clapp94}), SK94
(\cite{netterfield94}), and SP94 (\cite{gundersen94}) all report
detections of
anisotropy near this angular scale.

Quantifying CMBR anisotropy at the level of these detections
is an extremely challenging observational task (\cite{wilkinson95}).
Many potential systematic errors cannot be
unequivocally ruled out at the necessary levels, with the result
that any single observation cannot prudently be accepted without
an independent confirmation.
The results in this paper are our attempt to confirm the results
of our previous work.
By observing the same region of the sky with a second balloon flight,
we demonstrate the repeatability of our measurements in the 
presence of potential atmospheric noise and contamination from Earthshine.

We have reported earlier (\cite{cheng94}, hereafter Paper~I) our
observations of anisotropy of the CMBR from the first flight of MSAM
in 1992.
Our results from those observations were 1) a
positive detection of anisotropy, with the caveat that we could not
rule out foreground contamination by bremsstrahlung;
2) the identification of two particular
bright spots that were consistent with being unresolved sources.
This paper reports our first results from the 1994 flight of MSAM, which
observed an overlapping field.

\section{Instrument Description}

This instrument has been briefly described in Paper~I;
we give only an overview here.
It has four spectral bands at 5.6, 9.0, 16.5, and
22.5~\icm, giving sensitivity to CMBR and Galactic dust.  The
off-axis Cassegrain
telescope forms a 30\arcmin\ beam on the sky.  The chopping secondary
mirror moves this beam in a step motion 40\arcmin\ left and right of center.
The beam moves center, left, center, right with a
period of 0.5~s.  The detectors are sampled at 32~Hz, synchronously with
the chop.

The telescope is mounted on a stabilized balloon-borne platform.
The absolute pointing reference is provided by a star camera; positions between
camera fixes are interpolated using a gyroscope.
The telescope is shielded with aluminized panels so that the dewar feed horn, 
the secondary and most of the primary have 
no direct view of the Earth.

The gondola superstructure was changed between the 1992 and 1994
flights.  The previous superstructure as viewed from the telescope
had a substantial cross-section
of reflective material; in spite of
our efforts to shield it we were concerned about the telescope
being illuminated by reflected Earthshine.  The new
design is a cable suspension with considerably lower cross section
above the telescope.
Ground measurements indicate that rejection of signals from sources 
near the horizon is better than 75~dB in our longest wavelength
channel.

\section{Observations} 

The package was launched from Palestine, Texas at 00:59~UT 2~June~1994,
and reached its float altitude of 39.5~km at about 03:25~UT.
Science observations
ended with sunrise on the package at 12:04~UT.  During the flight
we observed Jupiter to calibrate the instrument and map the
telescope beam, scanned M31 (which will be reported in a future {\sl
Letter\/}), and integrated on the same CMBR field observed during the
1992 flight for 3.5 hours.

The CMBR observations were made as described in Paper~I.
The telescope observes near the meridian 8\arcdeg\ above the north
celestial
pole, and scans in azimuth $\pm 45\arcmin$ with a period of 1
minute.  The
scan is initially centered on a point 21\arcmin\ to the east of
meridian.  We track to keep this point centered in our scan until
it is 21\arcmin\ to the west of meridian, then jog 42\arcmin\ to the
east.  Each scan takes about 20 minutes, and half of each scan
overlaps the preceding scan.
We completed
4.5 such scans from 05:12 to 06:38~UT (we call
this section 1 of the data), and
completed an additional 7 scans from 07:22 to 09:43~UT (section 2).  The
observed field is two strips at declination $81\fdg8 \pm 0\fdg1$,
from right ascension 15\fh27 to 16\fh84,
and from 17\fh57 to 19\fh71 (all coordinates are J1994.5).
Fig.~\ref{f_fields} shows the
fields observed in the 1992 and 1994 flights.  
The overlap between the fields is better than half a beamwidth
throughout the flight.
Our ability to observe exactly the same position on the sky is
currently limited by the error in determining the position of the IR
beam center during the initial in-flight calibration, i.e., our
real-time determination of pointing is not as accurate as our post-flight
determination.

\section{Data Analysis}

The signal from the detectors is contaminated by spikes induced by cosmic rays
striking the detectors; we remove these spikes.
The data are calibrated by our observation of Jupiter.
The absolute pointing is
determined from star camera images.  The detector data
are analyzed to provide measurements of brightness in our four spectral
channels as a function of sky position.  These are then fit to a
spectral model to produce measurements of CMBR anisotropy and dust
optical depth.  These analyses and their results are described
in the following sections.

\subsection{Pointing}

We determined the pointing by matching star camera images
against a star catalog.  This fixes the position of the camera frame
at the time the exposure was taken.  Between exposures, position is
interpolated with the gyroscope outputs plus a small linear correction to
make the gyroscope readings consistent with the camera fixes.  This
correction is typically 2\arcmin\ in 20~minutes.  The relative
orientation of the camera frame and the IR telescope beam is fixed by
a simultaneous observation of Jupiter with the camera and the IR
telescope.  The resulting absolute pointing is accurate to 2\farcm5,
limited by the gyroscope drift correction.  The pointing analysis
was done in an
identical way for the 1992 flight, and has similar accuracy.

\subsection{Detector Data Reduction}

The instrument is calibrated by in-flight observations of Jupiter. The
brightness temperatures of Jupiter for our four spectral channels are
172, 170, 148, and 148~K, derived from the spectrum of Jupiter
observed by \cite{griffin86}. The apparent diameter of Jupiter during the
1994 flight is 42\arcsec. The uncertainty in the absolute
calibration is 10\%,
dominated by uncertainty in the antenna temperature of Jupiter.  The relative
calibration uncertainty between the 1992 and 1994 flights is 5\%,
due to noise in the observations of Jupiter.

The detector signal contains spikes, at a rate of 0.25--0.5~${\rm s}^{-1}$,
consistent with the hypothesis that they are due to cosmic
rays striking the detectors (\cite{charakhchyan78}), and with the rate
reported in Paper~I.  Cosmic rays deliver an
unresolved energy impulse to the detector; we remove them by fitting
the data to the impulse response function of the
detector/amplifier/filter chain.  
We give here our results for the 5.6~\icm\ channel; the numbers for the
other channels are similar.  Candidate spike locations are identified
using a $1.5\,\sigma$ threshold.  The data within 1~s (5 detector time
constants) are fit to a model of the response function.  About 2\% of
the spikes require a second spike 2--10 samples separated from the
first to be added to the fit.  If the resulting spike amplitude has
less than $3\,\sigma$ significance, the data is left as-is.  If the
fit is good, and the spike amplitude has more than $3\,\sigma$
significance, the spike template is subtracted.  5065 spikes are
subtracted out of 504,000 time samples.  (We allow either positive or
negative amplitudes; 90\% of the spikes have positive amplitude.)  If
the fit is poor, and the spike amplitude is significant, full data
records (64 samples, or 2 sec) before and after the spike are deleted.
317 spikes were eliminated this way, removing a total of about 6\% of
the data.

We estimate the instrument noise by measuring the variance in the
demodulated, deglitched data after removing a slow drift in time and
the mean in each sky bin.  This estimate is made for each 20~minute
segment of data, and is then propagated throught the remaining
processing.  All $\chi^2$ reported below are with respect to this
error estimate.

We divide the sky into bins that are small compared to the beamsize.
The bins are 0\fh057 in right ascension and 0\fdg12 in declination.
Due to sky rotation, the data
also need to be divided by angular orientation of the beam throw on
the sky; the bin size for this coordinate is 10\arcdeg.  The data are then
fit to a signal in each sky bin plus a model of long-term drift formed
from a cubic spline with knots every 12 minutes (2.5 minutes for the
16.5~\icm\ channel), plus terms for
gondola inclination, roll, and air pressure.  The simultaneous fit of long-term
drift and sky signal ensures that this fit does not bias our
observations of the sky.  This fit is done separately on each channel
and section of the flight.  The
resulting sky signals have bin-to-bin correlation, and we propagate a full
covariance matrix through the remainder of the analysis.
Sky bins containing less than 4~s of integration are deleted.
So that our error estimate, described in the preceding
paragraph, is unbiased by sky signal, we form the estimate from the
residuals of this fit, and iterate to obtain a consistent solution.

The data are demodulated in two different ways.
The double difference demodulation corresponds to 
summing the periods when the secondary is in the central position, and 
subtracting the periods when it is to either side.
This demodulation is least sensitive to 
atmospheric gradients and gondola swinging.
The single difference demodulation is formed by
differencing the period when the secondary is to the right from that when 
it is to the left, and ignoring the periods when the secondary is in the 
center.  We use the scan 
over Jupiter to deduce optimal demodulations of the infrared signal.

The 
binned dataset contains 90\% of all the data originally taken, with an 
achieved sensitivity in each of the four channels of 240, 150, 80,
and 230~\uK~$\sqrt{\rm s}$ 
Rayleigh-Jeans. For channels 1 and 2 this is 490 and 850~\uK~$\sqrt{\rm 
s}$ CMBR.  The offsets in the demodulated data for the different
channels
and demodulations range from 1 to 6~mK~RJ, smaller than those
reported in Paper~I.

\subsection{Spectral Decomposition}

At each sky bin, we fit the four spectral channels to a
model consisting of a CMBR anisotropy plus emission from warm Galactic dust.
The results are not very sensitive to the parameters of the dust model;
we use a dust temperature of
20~K and an emissivity index of 1.5 (consistent with \cite{wright91}).
The fit is
done separately for the single and double difference demodulations.
The $\chi^2/$DOF for the fit is 408/430 (double difference) and
448/430 (single difference).

Fig.~\ref{f_dust} shows the resulting fitted dust optical depth at
22.5~\icm.  For clarity this figure has been binned more coarsely and
does not distinguish between points at slightly different declination
or chop orientation; our analyses, however,
do not ignore these details.  We
have fit our observations to the \IRAS\ Sky Survey Atlas at 100~\micron\
(\cite{wheelock93}) convolved with our beam patterns, with
amplitude and offset as free parameters.  The resulting fit is
superimposed on Fig.~\ref{f_dust}.  The $\chi^2/$DOF of this fit is
262/210 for the double difference demodulation and 310/210 for the
single difference.  The ratio of optical depths between IRAS and our
data is consistent with an average dust emissivity spectral index
between our bands and 100~\micron\ of $\alpha = 1.40 \pm 0.16$
(still assuming a dust temperature of 20~K).

Our measurements of CMBR anisotropy are plotted in Fig.~\ref{f_cmbr}.
Superimposed are the measurements from 1992.  As noted earlier, there
is non-negligible correlation between the error bars on different sky
bins.  In making Fig.~\ref{f_cmbr} we have fit out the two largest
eigenmodes of the covariance matrix, and used error bars formed from
the diagonal of the covariance matrix after removing the two largest
eigenmodes; the result is that the error bars shown in the figure can
be approximately treated as uncorrelated.
(This procedure is similar to that used in \cite{fixsen94b} for the
{\sl COBE\/}/FIRAS calibration.)
The data have also been binned more
coarsely, as in Fig.~\ref{f_dust}.  We stress that these
steps are taken only for producing representative figures;
in all quantitative analyses we use
the full dataset and the full covariance matrix.
We are in the process of calculating the correlation for the MSAM1-92 data;
the 1992 data plotted here are identical to those in Paper~I.

\subsection{CMBR Anisotropy}

To set limits on anisotropy in the CMBR, we assume Gaussian
fluctuations with a Gaussian-shaped correlation function.  We set 95\%
confidence level upper and lower bounds on the total rms fluctuation
over the sky $(\sqrt{C_0})$, assuming this correlation function with a
given correlation angle $\theta_c$.  The method used is described in
Paper~I, though we now use a full covariance matrix for the 
instrument noise on the observations.
The upper and lower bounds from these observations for the
single and double difference demodulations are shown in
Fig.~\ref{f_deltat}.  The bounds for the correlation angles at which
the two demodulations are most sensitive are summarized in
Table~\ref{t_deltat}, which also shows results for the two sections of
the flight separately.
The confidence intervals for both demodulations are consistent with
those in Paper~I.

\section{Conclusions}

We observed the same field in our 1992 and 1994 flights in order to
determine if the detected signal was due to sidelobe pickup,
atmospheric noise,  or
other systematic effects, or was in fact present in the sky.  While we
are still in the process of completing
a detailed quantitative comparison of the two
datasets, it is apparent that the double difference
CMBR anisotropy features reproduce quite well.
This encourages us to believe
that the signal we see
in the double difference is present on the sky, and that
contamination from atmosphere or sidelobes is small compared to the
sky signal.
The single difference CMBR signal does not appear to reproduce as well.
Pending
the completion of the more thorough comparison,
we cannot rule out contamination in the single difference channel.

In Paper~I we pointed out that the anisotropy we
observe could be due to diffuse Galactic bremsstrahlung.  This
possibility remains, and will be addressed by our MSAM2 experiment,
which will observe the same fields in five bands over 65--170~GHz.

In Paper~I we raised the possibility that the ``sources'' at R.A. 19~h and
15~h were either foreground sources of a previously unknown population,
or non-Gaussian CMBR fluctuations.  This speculation was prompted by
our belief that such features were inconsistent with Gaussian
statistics.  More careful analysis by us
and independently by \cite{kogut94} has indicated that
features like these are in fact consistent with a variety of plausible
correlation  functions.
Observations by \cite{church95} at 4.7~\icm\ rule out the source MSAM15$+$82
being more compact than 2\arcmin.
Therefore removal of these regions in studies of CMBR
anisotropy, as we recommended in Paper~I, are a biased edit
of the data, and we no longer recommend it.

Our current conclusion is that the double difference, whole flight numbers
in Table~\ref{t_deltat} are a reliable estimate of CMBR anisotropy in the
observed regions.
When we include the 10\% uncertainty in the calibration, the resulting
limits are
$\Delta T/T = 1.9^{+1.3}_{-0.7}\times 10^{-5}$ 
(90\% confidence interval)
for total rms fluctuations.  In the band power
estimation of (\cite{bond95}), this is $\langle {\cal C}_l \rangle_B 
= 2.1^{+1.5}_{-0.9}\times 10^{-10}$ ($1\,\sigma$ limits), with
$\langle l \rangle = 263$.

The CMBR anisotropy channel, Galactic dust channel, pointing,
covariance matrices, and beammaps are publicly available.  For more
information, read {\tt
ftp://cobi.gsfc.nasa.gov/pub/data/msam-jun94/README.tex}.

\acknowledgments

We would like to thank the staff of the National Scientific Balloon
Facility, who
remain our willing partners in taking the calculated risks that result in
extremely successful flights.
W.~Folz and J.~Jewell traveled with us to the NSBF
to help with flight preparations.
T.~Chen assisted in building and testing our new star camera system.
We are grateful to M.~Devlin and S.~Tanaka for providing cappuccino at
the crucial moment in Palestine.
The Free Software Foundation provided the
cross-development system for one of the flight computers.
This research was supported by the NASA Office of Space Science,
Astrophysics Division.

\clearpage

\begin{deluxetable}{rccrrcrr}
\tablecolumns{8}
\tablecaption{Upper and lower bounds on total rms CMBR anisotropy ($\protect\sqrt{C_0}$) 
\label{t_deltat} }
\tablehead{
\colhead{} & \colhead{} & \multicolumn{3}{c}{MSAM1-94} &
\multicolumn{3}{c}{MSAM1-92} \\
\cline{3-5}\cline{6-8}
\colhead{} &	\colhead{} &
	\colhead{} & \colhead{Upper} & \colhead{Lower} &
	\colhead{} & \colhead{Upper} & \colhead{Lower}	\\
\colhead{$\theta_c$} &	\colhead{Section} &
	\colhead{R.A.} & \colhead{Bound} & \colhead{Bound} &
	\colhead{R.A.} & \colhead{Bound} & \colhead{Bound}	\\
\colhead{}	&	\colhead{} &
	\colhead{(h)} &	\colhead{(\uK)} &	\colhead{(\uK)}	&
	\colhead{(h)} &	\colhead{(\uK)} &	\colhead{(\uK)}	}

\startdata

\cutinhead{Single Difference}

0\fdg5 &	1 &	15.27--16.84 & 163 & 40 \nl
&		2 &	17.57--19.71 &  75 & 17 \nl
&		All &	15.27--19.71 &  79 & 30 & 14.44--20.33 & 116 & 53 \nl

\cutinhead{Double Difference}

0\fdg3 &	1 &	15.27--16.84 & 132 & 44 \nl
&		2 &	17.57--19.71 &  74 & 24 \nl
&		All &	15.27--19.71 &  78 & 34 & 14.44--20.33 &  97 & 50\nl
\enddata
\tablecomments{The limits in this table do not include the calibration
uncertainty.}
\end{deluxetable}

\clearpage
\bibliographystyle{aas}
\bibliography{cmbr}

\clearpage

\figcaption{The hatched circles show the sky coverage for the 1992 and
1994 flights as derived from the sky binning procedure.  The beam
size and chop spacing are indicated in the legend on the top of the plot.
The region shown covers R.~A. 14\fh0 to 20\fh5 and declination
$+80\arcdeg$ to $+84\arcdeg$.  The average declination difference
between the two flights is 10\arcmin. \label{f_fields} }

\figcaption{Dust optical depth $\times 10^6$ at 22.5~\icm.  The line
is the brightness expected from \IRAS\ 100~\micron\ data, with the
magnitude scaled to fit our observations. 
Scale at right is dust antenna temperature at 22.5~\icm.
a) double difference, b)
single difference. \label{f_dust} }

\figcaption{Measured CMBR anisotropy.  Points with diamonds are 1994
flight, crosses are 1992 flight.  
The telescope beam is superimposed.
a) double difference, b) single
difference. \label{f_cmbr} }

\figcaption{Upper and lower limits on total rms $\Delta T/T$ as a
function of correlation length for Gaussian-shaped correlation
functions.  Plotted are 95\% CL upper limits for the double difference
(solid), and single difference (long dashed); and 95\% lower limits
for the double difference (dashed) and single difference
(dotted). \label{f_deltat} }

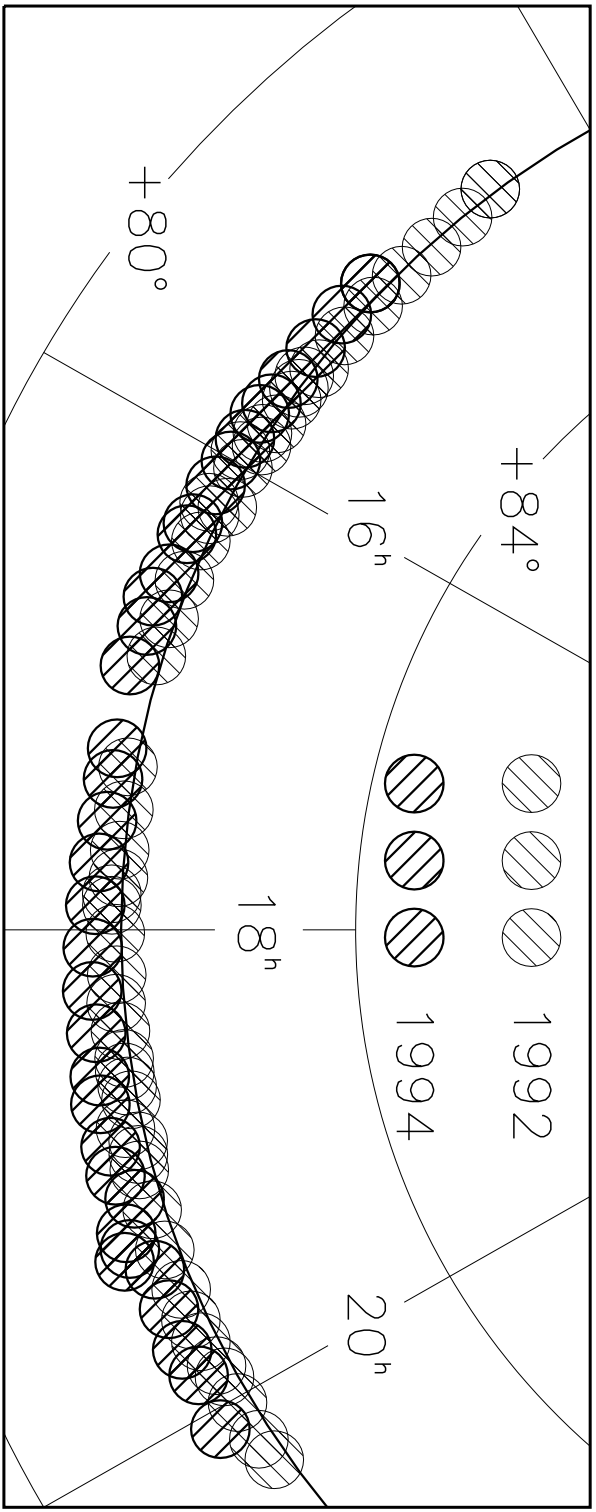
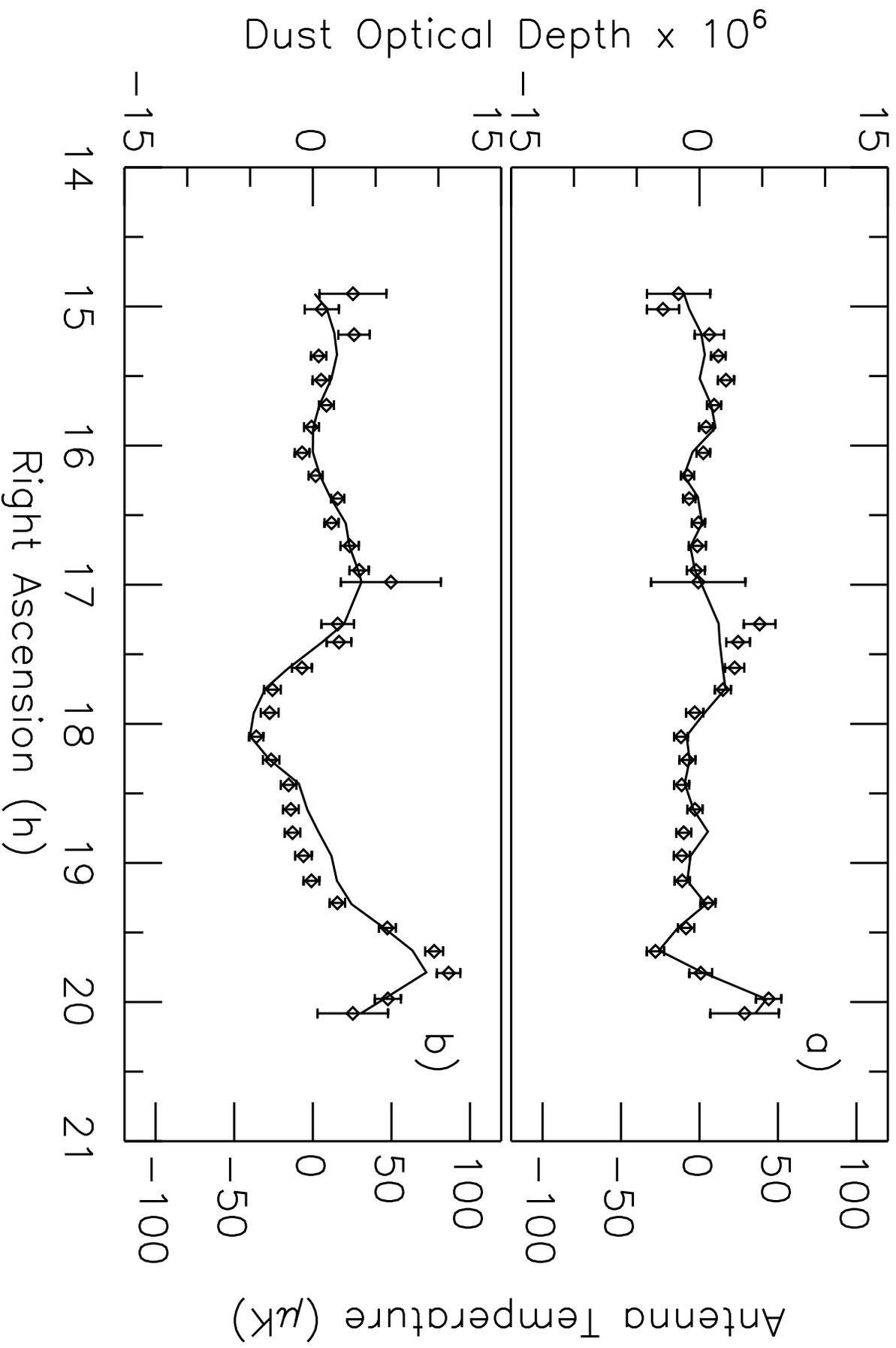
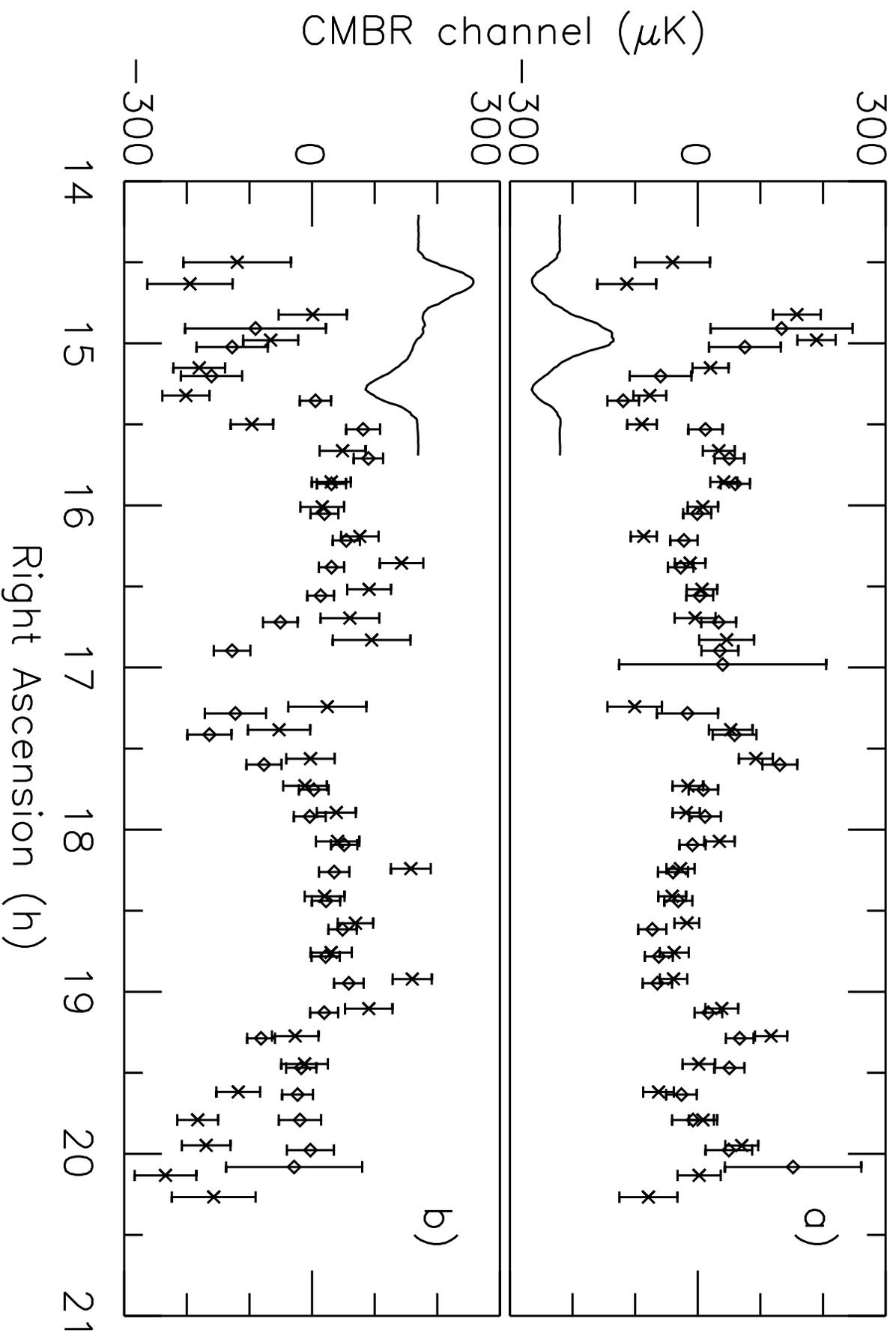
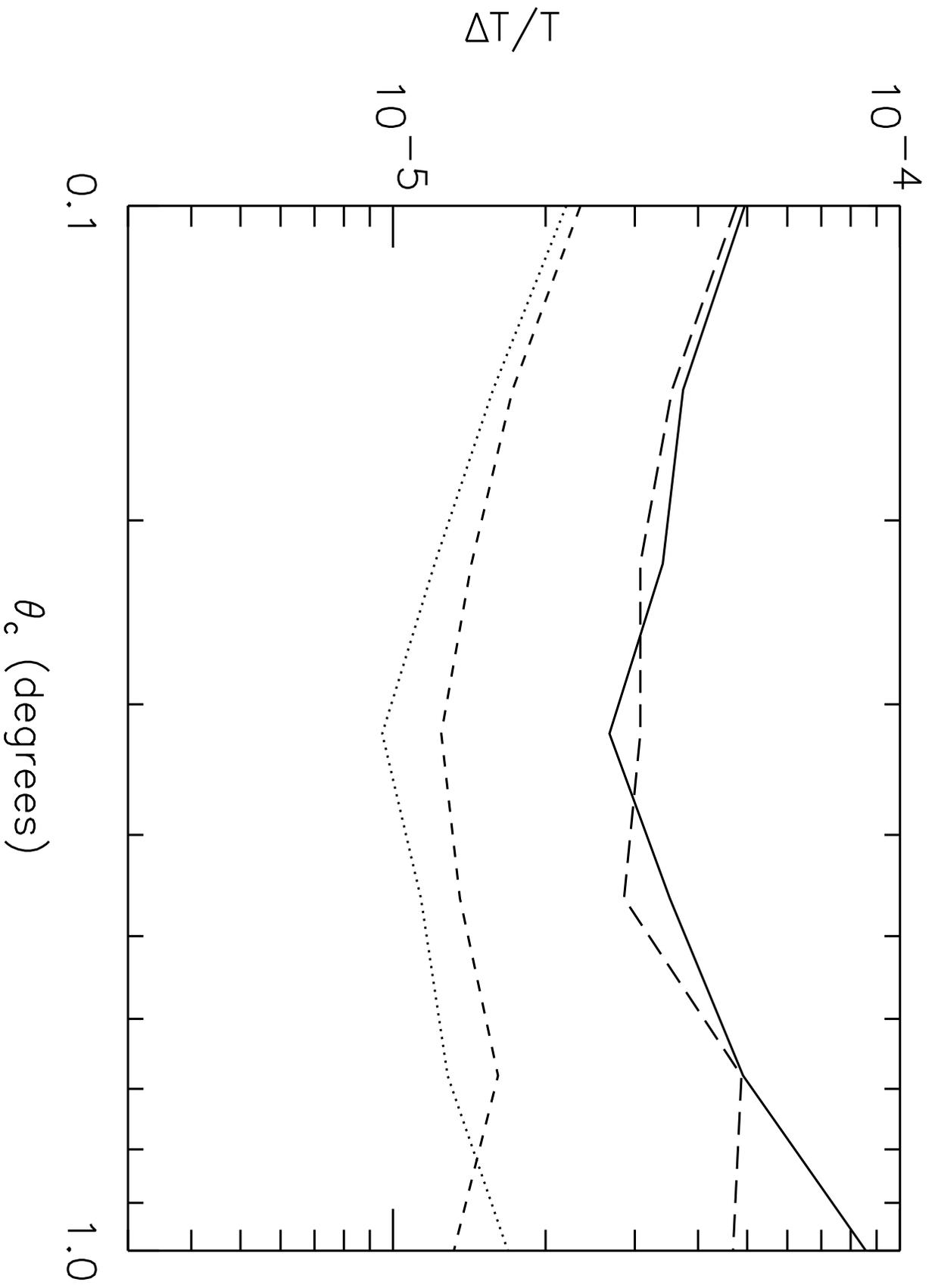
\end{document}